\documentclass[preprint]{revtex4-1}
\usepackage{newtxtext,newtxmath}

\usepackage{bm}
\usepackage{graphicx}
\usepackage{color}
\usepackage{hyperref}
\usepackage{amsmath}

\usepackage{color}

\usepackage[italicdiff]{physics}

\newcommand{\Order}[1]{\mathcal{O}\pqty{#1}}

\newcommand{\oH}{\hat{H}}
\newcommand{\oJ}[1]{\hat{J}^{(#1)}}
\newcommand{\oJp}{\hat{J}^{+}}
\newcommand{\oJm}{\hat{J}^{-}}
\newcommand{\oPi}{\hat{\Pi}}
\newcommand{\oa}{\hat{a}}
\newcommand{\ob}{\hat{b}}

\newcommand{\oR}{\hat{R}}
\newcommand{\oS}{\hat{S}}

\newcommand{\vk}{{\boldsymbol{k}}}

\newcommand{\dotot}{\hat{\rho}_{\mathrm{T}}}
\newcommand{\dos}{\hat{\rho}}
\newcommand{\dor}{\hat{\rho}_\mathrm{R}}
\newcommand{\dosl}{\tilde{\rho}}

\newcommand{\omegaa}{\omega_\mathrm{a}}
\newcommand{\omegac}{\omega_\mathrm{c}}
\newcommand{\cutoff}{\Lambda}

\newif\ifSHOWCHANGES
\usepackage{xcolor}
\SHOWCHANGEStrue 
\newcount\ChangesCount
\ifSHOWCHANGES
\long\def\REMOVEDSTRING#1{{\color{gray}{\it #1}}}

\def\IncChangesCount{\advance\ChangesCount by 1}
\long\def\REMOVED#1{\IncChangesCount\REMOVEDSTRING{#1}\relax}

\else
\long\def\REMOVED#1{\relax}

\fi

\begin{document}

\title{Dynamical properties of the finite-size Dicke model coupled to a thermal reservoir}

\author{R. Imai}
\email{rimai@asagi.waseda.jp}
\affiliation{Department of Electronic and Physical Systems, Waseda University, Tokyo 169-8555, Japan}

\author{Y. Yamanaka}
\email{yamanaka@waseda.jp}
\affiliation{Department of Electronic and Physical Systems, Waseda University, Tokyo 169-8555, Japan}

\date{October 6, 2018 by Y. Yamanaka, ver.2}
\date{October 11, 2018 by R. Imai, ver.3}
\date{October 12, 2018 by Editage, ver.3a}
\date{October 13, 2018 by R. Imai, ver.3b}
\date{October 13, 2018 by Y. Yamanaka, ver.4}
\date{\today}

\begin{abstract}
	We investigate the dynamical properties of the finite-size Dicke model 
	coupled to a photon reservoir in the dispersive regime.
	The system--reservoir coupling in our Hamiltonian
	includes counter-rotating terms, which are
	relevant in the strong atom--cavity coupling.
	Because the dispersive regime is considered, 
	the dynamics of low-energy states are described
	sufficiently accurately within the finite-dimensional subspace of the dressed states.
	Using the separation of the time scales between the system and the reservoir, we derive the Markovian quantum master equation in the subspace without ignoring the counter-rotating terms.
	The temporal evolution of the expectation of the cavity mode shows that the bifurcation of the long-lived state corresponds to the superradiant transition in the isolated model. The master equation explicitly gives the steady state solution. 
	The numerical results for the first-order correlation function on the steady state indicate that the strong atom--cavity coupling enhances the coherence and softens the dephasing in the superradiant region.
\end{abstract}

\maketitle

\section{Introduction}
Over recent decades, the characteristics of open quantum systems have been investigated in many fields of physics.
Alongside the progress in experimental techniques, further theoretical understanding of open systems is important.
In particular, novel quantum systems such as cold atoms and circuit quantum electrodynamics (QED) naturally include dissipation in varying degrees
\cite{Baumann2010-of,Brennecke2013}, and dissipation
can be artificially induced \cite{Tomita2017,Haeberlein2015}.
Moreover, in experiments on these systems, strongly correlated situations 
have been realized \cite{Baumann2010-of,Brennecke2013,Niemczyk2010,Yoshihara2016,Bloch2008}.
Thus, the physics of open quantum systems with strong interactions is attracting attention.

In cavity and circuit QED, in order to formulate dissipative effects, such as photon leakages, a reservoir is introduced that is coupled to the system.
In the standard procedure for deriving a quantum master equation (QME), the counter-rotating terms in the system--reservoir coupling, such as $\hat{a} \hat{b}$ or $\hat{a}^\dagger \hat{b}^\dagger$, are suppressed at the level of the Hamiltonian, for the reason that they
oscillate faster than the other terms. This approach is called the rotating-wave approximation (RWA) \cite{Breuer2007openquantum,QuamtumMeasurementAndControl2009}.
In recent years, it has pointed out that if a strong atom--cavity interaction is present, the RWA in the bare basis specified by the free Hamiltonian fails
to describe the physics of a dissipative system \cite{Beaudoin2011,Bamba2014}.
To overcome this failure, one can employ the RWA in a basis of dressed states or
exact eigenstates of the Hamiltonian of the system \cite{Beaudoin2011}.
In general, it is difficult to obtain the eigenstates of a strongly interacting system, 
and consequently to effectively understand open systems with strong interactions. 

In this article, we investigate the dynamical properties of the finite-size Dicke model \cite{Dicke1954,Hepp1973-zw,Wang1973}, which consists of a finite number of two-level atoms interacting with a single cavity mode coupled to a photon reservoir, where we keep the counter-rotating terms. 
The Holstein--Primakoff transformation \cite{Emary2003-vy}, 
which is valid in the thermodynamic limit, has been broadly employed for studies of the Dicke model.
Then, it has been observed that the Dicke model exhibits a transition from the normal phase
to the superradiant phase when the coupling strength between the atoms and the cavity mode reaches a critical value \cite{Emary2003-vy,Hepp1973-zw,Wang1973}.
The bosonic operators representing the excitations in the two phases differ from one another, and are determined 
separately \cite{Emary2003-vy}.
When the number of atoms is finite, the Holstein--Primakoff description 
involves highly non-linear correction terms in the Hamiltonian.
Thus, whereas the number of atoms in experimental situations is finite, 
the theoretical understanding of the model with a finite number remains limited.
There have been many previous studies addressing dissipation effects in
the Dicke model \cite{Nagy2011,Dalla2013,Brennecke2013,Fuchs2016}, 
where the Lindblad-type dissipater of bare operators
has mainly been employed. The existence of the Lindblad-type dissipater
shifts the transition point from that in the isolated case \cite{Dalla2013,Brennecke2013}, and a bifurcation in the stationary state that corresponds to the equilibrium superradiant transition is present in the semi-classical approximation \cite{Brennecke2013,Nagy2011,Oztop2012}.
The Lindblad-type dissipater is useful for a phenomenological description.
However, because it implicitly employs the RWA in the bare basis
\cite{Breuer2007openquantum,QuamtumMeasurementAndControl2009}, 
arguments using the Lindblad-type dissipater must be reconsidered 
in an interacting system, as mentioned above.
Specifically, near the phase boundary the energy gap is close to zero \cite{Larson2017}, 
and  
an approximation using a finite energy difference cannot be justified. 
It remains unclear how the phase transition,
originally defined in the thermodynamic limit of the isolated system, behaves
for the finite-size model with system--reservoir coupling that contains the counter-rotating terms. 

The outline of this work is as follows. 
In Sec.~\ref{sec:model}, we introduce the Dicke system coupled to a photon reservoir. The counter-rotating terms in the system--reservoir coupling are not ignored.
We consider the dispersive regime, in which the atomic level spacing is non-zero but very small in comparison with the cavity frequency. Then, the dynamics 
can only be accurately analyzed approximately in a low-energy subspace of the Hilbert space.
In Sec.~\ref{sec:QME}, we carefully derive the Markovian quantum master equation, assuming that the spectral density of the reservoir takes an ohmic form, and that the reservoir temperature is within a range determined by the parameters of the system Hamiltonian.
In Sec.~\ref{subsec:diss_dynamics}, we numerically solve the quantum master equation, and obtain the temporal evolution of the order parameter of the superradiant phase to capture the dynamical nature of the dissipative Dicke model. 
To verify the persistence of the coherence of the cavity mode, we also calculate the first-order correlation function that can be measure experimentally in Sec.~\ref{subsec:g1}.
Section \ref{sec:conclusion} is devoted to a summary, with some future outlooks.

\section{Dicke model in a dispersive regime}
\label{sec:model}
A system that consists of $N$ atoms and a single cavity mode is modeled by the Dicke Hamiltonian.
Let us consider the Dicke model with a $D$-dimensional photon reservoir.
In this study, the Hamiltonian of the total system is assumed as
\begin{align}
\oH_\mathrm{T} = \oH_\mathrm{D} + \oH_\mathrm{R} + \oH_\mathrm{DR} \,,
\end{align}
where
\begin{align}
\oH_\mathrm{D} &= \omegaa \oJ{3} + \omegac\oa^\dagger \oa + \frac{\lambda}{\sqrt{N}} (\oa + \oa^\dagger) (\oJp + \oJm) \,, \\
\oH_\mathrm{R} &= \int d^D \vk \, \Omega_\vk \ob_\vk^\dagger \ob_\vk \,, \\
\oH_\mathrm{DR} &= \oS \oR\,,
\end{align}
with
\begin{align}
\oS &= \oa + \oa^\dagger\,, \label{eq:def_S} &
\oR &= \int d^D \vk\, g(\Omega_\vk) (\ob_\vk + \ob_\vk^\dagger) \,.
\end{align}
Here, $\oa$ and $\ob_\vk$ denote the bosonic annihilation operators for the cavity mode with the frequency $\omegac$ and a reservoir mode with the wave vector $\vk$, respectively. The operators $\oJ{i}$ ($i=1,2,3$) are pseudospin operators for the collection of $N\,(=2J\,)$ two-level atoms, whose level splitting is given by $\omegaa$. Furthermore,
$\hat{J}^\pm$ are the ladder operators for the pseudospin, which are defined by $\oJ{1} \pm i \oJ{2}$.
Note that the coupling term between the Dicke system and the reservoir includes the counter-rotating terms, such as $\oa \ob_\vk$ and $\oa^\dagger \ob_\vk^\dagger$.
For the reservoir, we assume that the dispersion relation is linear, namely $\Omega_\vk = \abs{\vk}$, and that the coupling coefficient to the system $g(\omega)$ is real and depends only on the frequency of the mode.

In this study, we suppose that the level spacing of atoms $\omegaa$ is
considerably smaller than the cavity frequency $\omegac$.
This is known as a dispersive regime, where coherent
excitations of two-level atoms play a leading role in the low-energy physics, as we will see.
Then, we decompose the Hamiltonian of the Dicke system $\oH_\mathrm{D}$ into two parts:
\begin{align}
\oH_0 &= \omegac \pqty{\oa^\dagger + \frac{2 \lambda}{\sqrt{N} \omegac} \oJ{1}}\pqty{\oa + \frac{2 \lambda}{\sqrt{N} \omegac} \oJ{1}} \,, \\
\oH_1 &=  - \frac{4 \lambda^2}{N \omegac} (\oJ{1})^2 +  \omegaa \oJ{3}\,,
\end{align}
with $\oJ{1} = (\oJp + \oJm)/2$.
Because $\oH_0$ represents a quadratic form of the displaced bosonic operator, we can obtain all of the exact eigenstates of $\oH_0$ as
\begin{align}
\ket{\phi_{\mu,n}} = \ket{\mu} \otimes D_\mu \ket{n}\,,
\end{align}
where $\oJ{1} \ket{\mu} = \mu \ket{\mu}$, $\oa^\dagger \oa \ket{n} = n \ket{n}$, 
and the displacement operator $D_\mu$ is defined by 
\begin{align}
D_\mu = \exp{\alpha_\mu \oa^\dagger - \alpha_\mu^\ast \oa}
\quad \text{with} \quad \alpha_\mu = -\frac{2 \lambda \mu}{\sqrt{N} \omegac} \,,
\label{eq:def_D_alpha}
\end{align}
which induces the transformation $D_\mu \oa D^\dagger_\mu = \oa - \alpha_\mu$.
The eigenenergy to which $\ket{\phi_{\mu,n}}$ belongs is independent of $\mu$,
and is given by $\epsilon_{n} = n \omegac$. For each $\epsilon_{n}$, there exist $2J+1$ 
degenerate eigenstates.
The largest magnitudes of the matrix elements of
the first and second terms in $\oH_1$ are estimated
as $N \lambda^2 / \omegac$ and $N \omegaa / 2$, respectively.
In this study, we assume that $\omegac$ is significantly larger than these magnitudes, or
\begin{align}
	\sqrt{N} \lambda \ll \omega_c ~ \land ~
	\frac{N}{2} \omega_a \ll  \omega_c \,.
	\label{eq:dispersive_cond}
\end{align}
This situation was considered in Ref.~\cite{Barberena2017}, 
and was called the photon-number-dependent 
regime, as a restricted case of the dispersive regime.
The excited eigenstate of the displaced mode $\ket{\phi_{\mu,n}}$ for $n > 1$ 
has an eigenenergy larger than $\omegac$.
Thus, these states oscillate rapidly, and can be ignored when we are 
interested in the low-energy or low-temperature dynamics.
Thus, the low-energy dynamics can be well described in the
$2J+1$-dimensional subspace spanned by the displaced 
vacua $\ket{\phi_{\mu,0}}$ for $\mu = -J, -J+1, \dots, J$, which 
simplifies our analysis of
the Dicke model both in the normal and superradiant phases.
The dynamics in this subspace
is governed by $\oH_1$, which is equivalent to the Lipkin--Meshkov--Glick model \cite{Barberena2017}.

\section{Derivation of QME}
\label{sec:QME}
Here, we derive a QME for the dissipative Dicke system, in a similar manner to the well-known derivation of the high-temperature master equation \cite{Breuer2007openquantum,Caldeira1983}. 
Our derivation includes a system--reservoir interaction without the RWA.
Nevertheless, the QME takes a Markovian form under the assumption that the reservoir temperature lies within a certain range determined by the separation of the energy scales between the system and the reservoir.
The Markovian form permits us to analyze the QMEnumerically without difficulty, as in Sec.~\ref{sec:numerical}.

We now decompose the Hamiltonian of the total system into two parts:
\begin{align}
\oH_\mathrm{U} = \oH_0 + \oH_\mathrm{R} \,, \qquad \oH_\mathrm{Int} = \oH_1 + \oH_\mathrm{DR}\,.
\end{align}
Let us move to the interaction picture, where the time-dependence of an
operator $O$ is defined as $O(t) = e^{i \oH_\mathrm{U} t} O e^{-i \oH_\mathrm{U} t}$.

The temporal evolution of the density operator for the total system $\dotot(t)$ is described by the von Neumann equation:
\begin{align}
\dv{t} \dotot(t) &= -i \comm{\oH_\mathrm{Int}(t)}{\dotot(t)}
\label{eq:vNeq1}
\\
&= -i \comm{\oH_1(t)}{\dotot(t)}
-i \comm{\oH_\mathrm{DR}(t)}{\dotot(t)}
\label{eq:vNeq2}
\end{align}
Integrating \eqref{eq:vNeq1} over $t$ from 0 to $t$, we obtain
\begin{align}
\dotot(t) = \dotot(0) -i \int_{0}^t ds \, 
\comm{\oH_\mathrm{Int}(s)}{\dotot(s)}.
\label{eq:vNeq_integrated}
\end{align}
Here, we assume that the initial state at $t=0$
is given in the product form of the density operators of the partial systems, namely
\begin{align}
\dotot(0) = \dos(0) \otimes \dor \,,
\end{align}
and that the reservoir is in a thermal equilibrium with the inverse temperature $\beta$, or $\dor \propto e^{-\beta \oH_R}$.
By substituting Eq.~\eqref{eq:vNeq_integrated} into the second term of Eq.~\eqref{eq:vNeq2}, we have
\begin{align}
\dv{t} \dotot(t) &= -i \comm{\oH_1(t)}{\dotot(t)}
- \int_0^t ds \, \comm{\oH_\mathrm{DR}(t)}{\comm{\oH_\mathrm{Int}(s)}{\dotot(s)}}
\label{eq:vNeq3}
\end{align}
We assume that the density operator for the total system is a simple product of the two density operators without correlation, namely
\begin{align}
\dotot (t) \approx \dos(t) \otimes \dor \,.
\end{align}
After taking a partial trace with respect to the reservoir states, Eq.~(\ref{eq:vNeq3}) becomes
\begin{gather}
\dv{t} \dos(t) = -i \comm{\oH_1(t)}{\dos(t)}
- \int_0^t ds \, \mathcal{L}_\mathrm{R}(t, t-s) \dos(t-s) \,,
\label{eq:QME_Born}
\end{gather}
where $\mathcal{L}_\mathrm{R}$ is a superoperator that originates 
from the interaction with the reservoir, and is defined as
\begin{align}
\mathcal{L}_\mathrm{R} (t_1,t_2) O = \Big[ \oS(t_1) \oS(t_2) O - \oS(t_2) O \oS(t_1) \Big] C(t_1-t_2) + \text{(h.c.)}\,.
\label{eq:def_L}
\end{align}
Here, the correlation function $C(t)$, which completely characterizes the reservoir, is defined by $C(t) = \tr_R \bqty{\oR(t) \oR(0) \dor}$.
This correlation function depends not only on the reservoir itself, but also on the system--reservoir coupling strength $g_k$, via the definition of $\oR$.
Using the spectral density of the reservoir $J(\omega)$, defined by
\begin{align}
J(\omega) = [g(\omega)]^2 \int d^D \vk \, \delta (\omega - \Omega_k)\,,
\end{align}
we can express $C(t)$ as
\begin{align}
C(t) = \int_0^{\infty} d\omega \, J(\omega) \bqty{2 n_\mathrm{B}(\omega) \cos(\omega t) + e^{-i \omega t}}\,,
\label{eq:C(t)}
\end{align}
where $n_\mathrm{B}(\omega)$ denotes the Bose--Einstein distribution function $n_\mathrm{B}(\omega) = [e^{\beta \omega} - 1]^{-1}$. The Fourier transform of $C(t)$, denoted by 
$\tilde{C}(\omega)$, is expressed as 
\begin{align}
\tilde{C}(\omega) = \frac{1}{2\pi} \int_{-\infty}^\infty dt\, C(t) e^{i \omega t}
= J(\abs{\omega}) \bqty{n_\textrm{B}(\abs{\omega}) + \theta(\omega)} 
\label{eq:tilde_C}
\end{align}
where $\theta(x)$ denotes the Heaviside step function.
Note that $C(t)$ has a symmetry such that $C(-t) = [C(t)]^\ast$, which forces $\tilde{C}(\omega)$ to be real-valued.

To simplify the temporal integral in the QME \eqref{eq:QME_Born}, we estimate
typical time-scales of the system and the reservoir.
For the system, as the largest energy gap in the low-energy sector is estimated as $\Delta \epsilon_\mathrm{max} = \max (N \lambda^2 / \omegac, N \omegaa /2)$ and
the inverse of this gap gives the fastest rate of change in $\dos(t)$, 
the typical short-time scale of the system $\tau_\mathrm{S}$ is
\begin{align}
\tau_\mathrm{S} \approx \frac{1}{\Delta \epsilon_\mathrm{max}} = \min \pqty{\frac{\omegac}{N \lambda^2}, \frac{2}{N \omegaa}} \,.
\end{align}
On the other hand, the typical time scale of the reservoir $\tau_\mathrm{R}$ depends on the spectral density.
For definiteness, we take a spectral density in the ohmic form, i.e.,
$J(\omega) = \eta \omega$, where the dimensionless parameter $\eta$ determines the strength of the dissipation \cite{Breuer2007openquantum,QuamtumMeasurementAndControl2009}.
The ohmic spectral density is realized when the system is coupled to
a one-dimensional reservoir with $g(\omega) \propto \sqrt{\omega}$
or a two-dimensional reservoir with a constant $g$.
Then, we analytically integrate Eq.~\eqref{eq:C(t)} to obtain
\begin{align}
C(t) = -\frac{\eta \pi^2}{\beta^2 \sinh^2 (\pi \abs{t} / \beta)}\,.
\label{eq:C(t)_ohmic}
\end{align}
The details of this manipulation are presented in App.~\ref{app:ohmic_cf}.
For $t \gg \beta$, we approximate Eq.~\eqref{eq:C(t)_ohmic} as
\[
C(t) \approx -\frac{\eta}{\tau_\mathrm{R}^2} \exp \pqty{-\frac{\abs{t}}{\tau_\mathrm{R}}}\,, 
\]
where $\tau_\mathrm{R} = {\beta}/{2 \pi}$ determines the typical time scale of the reservoir, and is determined by the reservoir temperature. 
The correlation function decays exponentially in time with the time scale $\tau_R$, 
which represents the duration in which the integrand is non-negligible in  the temporal 
integration Eq.~\eqref{eq:QME_Born}.
When $\tau_\mathrm{S}$ is significantly larger than $\tau_\mathrm{R}$, $\dos(s)$ 
becomes almost constant in the integration of Eq.~\eqref{eq:QME_Born}.
Therefore, $\dos(s)$ can be replaced by $\dos(t)$, which is known as the Markov approximation.
We find that the condition for this approximation, $\tau_\mathrm{R} \ll \tau_\mathrm{S}$, is equivalent to
\begin{align}
\max \pqty{\frac{N \lambda^2}{\omegac}, \frac{N \omegaa}{2}} \ll \frac{1}{\beta} \,,
\label{eq:beta_limit1}
\end{align}
which gives a lower limit on the reservoir temperature.
The suppression of excited states, discussed in Sec.~\ref{sec:model}, demands
an upper limit on the temperature of
\begin{align}
\frac{1}{\beta} \ll \omegac\,.
\label{eq:beta_limit2}
\end{align}
Combining the inequalities \eqref{eq:beta_limit1} and \eqref{eq:beta_limit2}, we have that
\begin{align}
\frac{1}{\omegac} \ll \beta \ll \min \pqty{\frac{\omegac}{N \lambda^2}, \frac{2}{N \omegaa}} \,. 
\label{eq:beta_region}
\end{align}
Under the condition \eqref{eq:beta_region}, we can utilize the Markov approximation to rewrite the integral of \eqref{eq:QME_Born} as
\begin{align}
\int_0^t ds \, \mathcal{L}_\mathrm{R} (t,t-s) \dos(t-s)
\to
\int_0^\infty ds \, \mathcal{L}_\mathrm{R} (t,t-s) \dos(t)
\label{eq:markov_approx}
\end{align}

As presented in Sec.~\ref{sec:model}, for the low-energy dynamics we can utilize the density operator in the lower-energy sector of the system, denoted 
by $\dosl$. This is expressed by the states $\ket{\phi_\mu} \equiv \ket{\phi_{\mu,0}}$ for $\mu = -J, -J+1,\dots, J$ as
\begin{align}
\dosl(t) = \sum_{\mu,\nu = -J}^{J} \dosl_{\mu,\nu}(t) \dyad{\phi_\mu}{\phi_\nu}\,,
\label{eq:expansion_density_op}
\end{align}
where $\dosl_{\mu,\nu}$ is an element of a Hermitian matrix, 
$\dosl_{\mu,\nu} = \dosl_{\nu,\mu}^\ast$.
Using Eqs.~\eqref{eq:QME_Born} and \eqref{eq:markov_approx}, as discussed in App.~\ref{app:Evaluation_diss}, we find that the temporal evolution of the matrix element of $\dosl(t)$ is represented as
\begin{align}
\dv{t} \dosl_{\mu,\nu} = -i\comm{H_1(t) + \Sigma}{\dosl}_{\mu,\nu} - [\Gamma^\text{ex.}   + \Gamma^\text{de.}_{\mu,\nu}] \dosl_{\mu,\nu}
\label{eq:QME_full}
\end{align}
where $\mathcal{O}_{\mu,\nu}$ denotes $\bra*{\phi_\mu} \mathcal{O} \ket*{\phi_\nu}$ for an operator
$\mathcal{O}$, and
\begin{align}
\Sigma_{\mu,\nu} &= - 16 \delta_{\mu,\nu} \frac{\lambda^2}{N \omegac^2} \mu^2 \mathrm{PV} \int d \omega \, \frac{\tilde{C}(\omega)}{\omega} \,, 
\label{eq:Sigma} \\
\Gamma^\text{ex.} &= 2 \tilde{C}(-\omegac) \,, 
\label{eq:Gamma_ex}\\
\Gamma^\text{de.}_{\mu,\nu} &= 16 \tilde{C}(0) \frac{\lambda^2}{N \omegac^2} (\mu - \nu)^2 \,.
\label{eq:Gamma_de}
\end{align}
Furthermore, $\mathrm{PV}$ denotes the Cauchy principal value.
In deriving Eq.~\eqref{eq:QME_full} we do not employ any additional approximations, such as the RWA.

The term $\Sigma_{\mu,\nu}$, which is diagonal with respect to the $\mu$-index,
can be renormalized to $\omega_c$ through 
an energy shift induced by the interaction with the reservoir.
Consider a sum of the diagonal matrix elements of $\oH_1$,
\begin{align}
\bra{\phi_\mu} \oH_1(t) \ket{\phi_\mu} = - \frac{4 \lambda^2 \mu^2}{N \omegac}
\,,
\end{align}
and $\Sigma_{\mu, \mu}$,
\begin{align}
\bra{\phi_\mu} \oH_1(t) \ket{\phi_\mu} + \Sigma_{\mu, \mu}
=-\frac{4 \lambda^2 \mu^2}{N\omega_c}
\left( 1-\frac{N\omega_c \Sigma_{\mu, \mu}}{4 \lambda^2 \mu^2} \right)
\approx  \frac{4 \lambda^2 \mu^2}{N(\omega_c+\delta \omega_c)}
\end{align}
with 
\begin{align}
\delta \omegac = 4 \mathrm{PV} \int d \omega \, \frac{\tilde{C}(\omega)}{\omega} \,.
\end{align}
The renormalized cavity frequency $\omegac+\delta \omegac$ is simply denoted by $\omegac$ in the following.

The term $\Gamma^\text{ex.}$ decreases the trace of $\dosl(t)$, or breaks the conservation of probability.
This decrease can be interpreted as the flow of the probability out of the low-energy sector, owing to reservoir-induced excitations.
This flow depends on the temperature of the reservoir via the relation $\tilde{C}(-\omegac) =  \eta \omegac n_B(\omegac)$.
In the range for $\beta$ in Eq.~\eqref{eq:beta_region}, the ratio 
$\tilde{C}(-\omegac)/\tilde{C}(0)$ is approximately 
$\beta\omegac \exp(-\beta\omegac)$, and is considerably small. This means 
that $\Gamma^\text{ex.}$ can be neglected in the presence of 
the $\Gamma^\text{de.}_{\mu,\nu}$ term in Eq.~(\ref{eq:Gamma_de}).

The term $\Gamma^\text{de.}_{\mu,\nu}$ causes
a dephasing effect in the temporal evolution, which is of significance in this study.
The coefficient $\tilde{C}(0)$ in $\Gamma^\text{de.}_{\mu,\nu}$
is evaluated as $\tilde{C}(0) = \eta/\beta$ for the ohmic spectral density.
Then, the degree of dephasing is controlled by the single parameter
\begin{align}
\gamma = \frac{16\eta}{\beta}\,.
\end{align}

In summary, the QME \eqref{eq:QME_full} is simplified as
\begin{align}
\dv{t} \dosl^\mathrm{S}_{\mu,\nu} (t) = [\mathcal{L} \dosl^\mathrm{S}]_{\mu,\nu}\,, 
\label{eq:QME_simplified_Schrodinger}
\end{align}
where $\mathcal{L}$ denotes the Liouvillian, defined by
\begin{align}
[\mathcal{L} \dosl^\mathrm{S}]_{\mu,\nu} = -i \comm{\oH_1}{\dosl^\mathrm{S}(t)}_{\mu,\nu}
- \gamma \frac{\lambda^2}{N \omegac^2} (\mu - \nu)^2 \dosl^\mathrm{S}_{\mu,\nu} (t)
\label{eq:def_L2}
\end{align}
in the Schr\"odinger picture, and the density matrix $\dosl^\mathrm{S}(t)$ 
is defined by
$\dosl^\mathrm{S}_{\mu,\nu}(t) = e^{-i \oH_0 t} \dosl_{\mu,\nu}(t) e^{i \oH_0 t}$.
While the first term in Eq.~\eqref{eq:def_L2} represents a 
unitary dynamical evolution, the second term, whose 
origin is the system--reservoir interaction, causes
non-unitary dephasing.
Because the right-hand side of Eq.~\eqref{eq:QME_simplified_Schrodinger} does not contain a time integral and $\mathcal{L}$ is independent of the time variable, the QME is Markovian, which facilitate the following numerical analysis.
The steady state that satisfies the condition
$\mathcal{L} \dosl^\mathrm{S,ss} = 0$ 
is found to be
\begin{align}
\dosl^\mathrm{S,ss}_{\mu,\nu} = (N+1)^{-1} \delta_{\mu,\nu}\,.
\label{eq:steadystate}
\end{align}
This steady state is maximally mixed, or highly thermal.
Note that the density operator ${\tilde \rho}^{S,ss}(t)$
corresponding to the matrix $\dosl_{\mu,\nu}$ in Eq.~\eqref{eq:steadystate}
depends on the parameters $\omegaa, \omegac, \lambda$ and $N$, because
the states in the low-energy subspace depend on them.

\section{Numerical results}
\label{sec:numerical}
\subsection{Dissipative dynamics of the cavity field}
\label{subsec:diss_dynamics}

To investigate the dynamics of the cavity mode interacting
with the reservoir, we numerically integrate the QME \eqref{eq:QME_simplified_Schrodinger}.
In the following, we fix the number of atoms as $N = 16$, except in Fig.~\ref{fig:lambda_tauc}, and the ratio of the cavity mode frequency to the atomic level spacing as $\omegac / \omegaa = 400$.
Then, because $N \omegaa / (2 \omegac)  = 0.02$, the second condition of Eq.~\eqref{eq:dispersive_cond} is automatically satisfied.
On the other hand, the first condition of Eq.~\eqref{eq:dispersive_cond} 
now yields an upper limit on the coupling strength, namely $\lambda \ll 100 \omegaa$. 
We will perform our analysis within this limitation.

\begin{figure}
	\centering
	\includegraphics[width=3.4in]{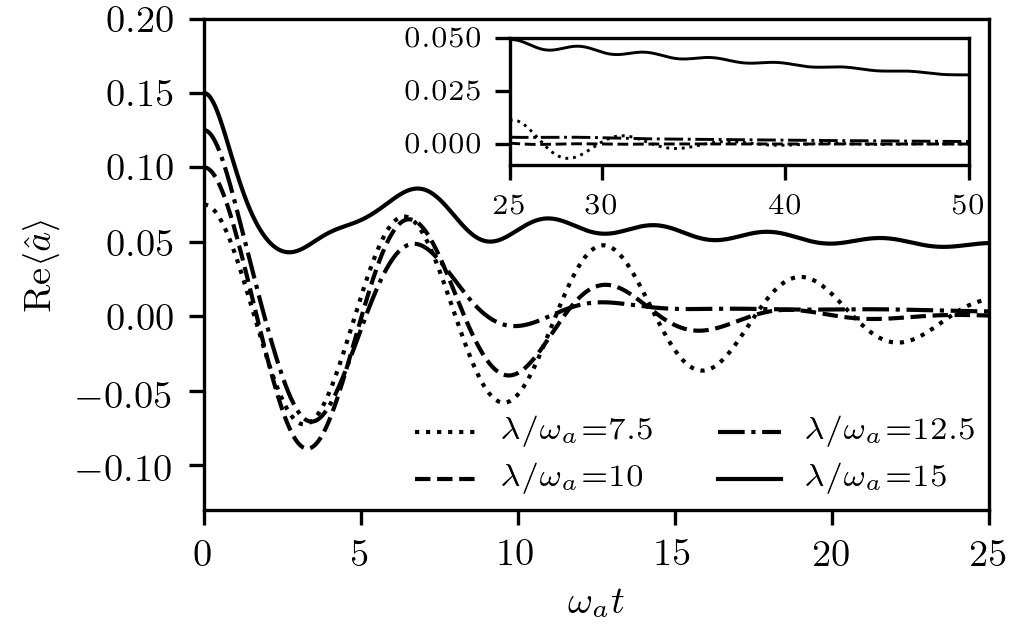}
	\caption{Temporal evolutions of $\Re \ev{\oa}$ for various values of the coupling strength $\lambda$ are shown.
		The inset shows the dynamics for $\omegaa t > 25$.
		The initial state is taken to be $\dyad{\phi_{-J}}$. 
		For all the plots, $N = 16$, $\omegac / \omegaa = 400$, and $\gamma/\omegaa = 100$.
	}
	\label{fig:QME_1}
\end{figure}

We first calculate the temporal evolution of the expectation 
$\Re \ev{\oa}$, because it is the order parameter that indicates 
the breaking of the $\mathrm{Z}_2$ symmetry on the isolated Dicke model, and this has a non-zero value in the superradiant phase.
The results are depicted in Fig.~\ref{fig:QME_1}.
Here, we choose an initial state $\dosl(0) = \dyad{\phi_{-J}}$,
where the cavity mode is perfectly coherent and the two-level atoms are collectively
excited. This is one of the localized ground states of 
the Dicke model with extremely strong coupling, or $\lambda \to +\infty$.
Thus, the choice of the initial condition corresponds
to the dynamics after a quench at $t = 0$, in such a manner that
the coupling strength suddenly changes from infinity to a finite value and 
the decoherence sets in simultaneously.
This quench setting helps us to capture the nature of the dynamics in the present dissipative model, although it may be difficult to realize in a real physical system.
When the coupling $\lambda$ is weak, $\Re \ev{\oa}$ exhibits an oscillation around
zero and approaches zero, as seen in Fig.~\ref{fig:QME_1}.
For $\lambda \gtrsim 12.5$, although a similar damped oscillation is observed, the center of the oscillation is shifted to a positive value, which becomes larger as 
$\lambda$ increases.
This implies that the system lies in a long-lived state
with non-vanishing $\Re \ev{\oa}>0$.
The transition point of this qualitative change cannot be clearly identified.
Instead, we observe that the behavior gradually changes at a coupling strength slightly larger than $\lambda_\mathrm{c} = \sqrt{\omegaa \omegac}/2 = 10 \omegaa$, which suggests 
that this change in behavior is closely related
to the superradiant phase transition.

If $\dosl^\mathrm{S}_{\mu, \nu}(t) = f_{\mu, \nu}(t)$ is 
a solution of the QME, then we readily observe that 
the inverted form $f_{-\mu, -\nu}(t)$ is also a solution.
This is clear from the fact that the unitary transformation
\begin{align}
\oPi = \exp \bqty{-i \pi \pqty{\oa^\dagger \oa + \oJ{3} + J}}\,
\end{align}
commutes with $\oH_1$, and satisfies the property
$\oPi \ket*{\phi_\mu} = \ket*{\phi_{-\mu}}$.
Therefore, when the initial state $\dyad{\phi_{-J}}$ is replaced by 
$\dyad{\phi_{+J}}$, $\Re \ev{\oa}$ only changes its sign, and
the long-lived state $\Re \ev{\oa} < 0$ exists for large $\lambda$.
This situation also corresponds to the breaking of $\mathrm{Z}_2$ symmetry in the superradiant phase \cite{Emary2003-vy}.

\subsection{First-order correlation function}
\label{subsec:g1}

We also calculate the normalized first-order correlation function \cite{Glauber1963} on the steady state, defined by
\begin{align}
g^{(1)} (t) = \frac{\ev{\oa^\dagger(t) \oa(0)}_\mathrm{ss}}{\sqrt{\ev{\oa^\dagger(t) \oa(t)}_\mathrm{ss} \ev{\oa^\dagger(0) \oa^(0)}_\mathrm{ss}}} \,,
\label{eq:def_g1}
\end{align}
where $\ev{\cdots}_\mathrm{ss}$ denotes an expectation in the steady state, namely $\ev{\cdots}_\mathrm{ss} = \Tr \bqty{\cdots \dosl^\mathrm{S,ss}}$. The function
$g^{(1)} (t)$, indicating the degree of possible interference 
of the cavity mode at two different times, can in principle 
be measured with a Mach--Zehnder interferometer in an optical system.
For instance, a method for circuit QED systems is presented in Ref.~\cite{Bozyigit2010}. 
Thus, $g^{(1)} (t)$ calculated below can be checked in future experiments when the 
steady state is realized.

\begin{figure}
	\centering
	\includegraphics[width=3.4in]{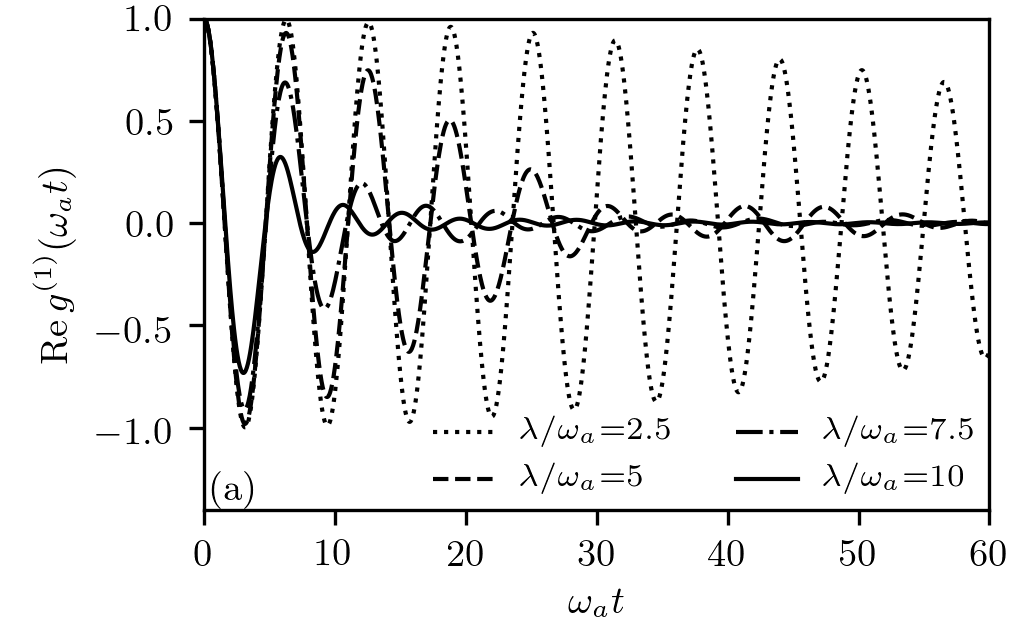}
	\\
	\includegraphics[width=3.4in]{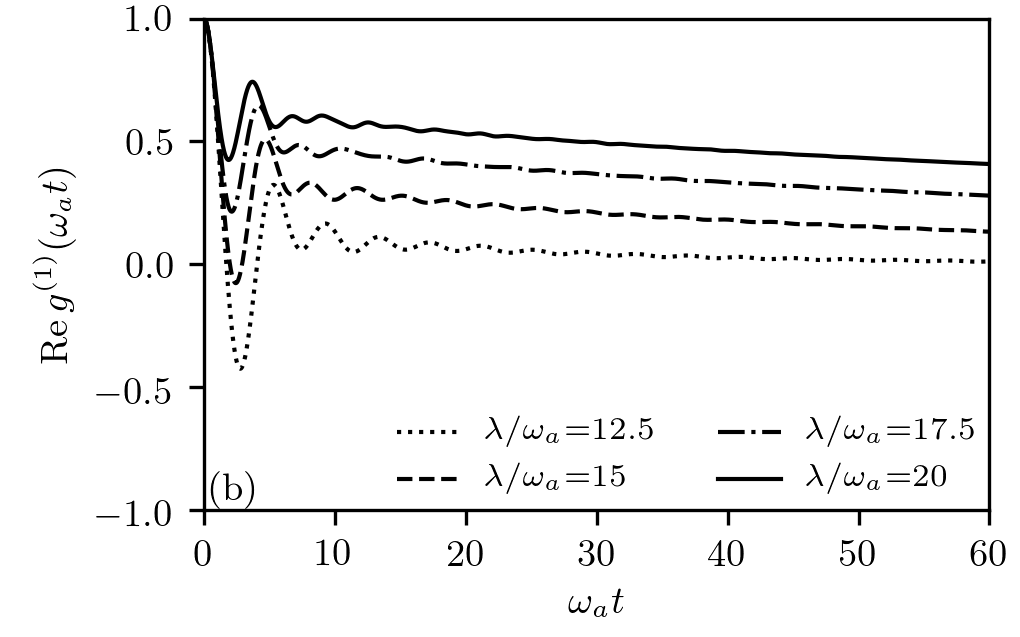}
	\caption{The first-order correlation function $\Re g^{(1)}$ for the various coupling values $\lambda / \omegaa = 2.5, 5, 7.5, 10$ in (a) and $\lambda / \omegaa = 12.5, 15, 17.5, 20$ in (b). The other parameters are fixed as $N = 16$, $\omegac / \omegaa = 400$, and $\gamma/\omegaa = 100$.}
	\label{fig:g1_1}
\end{figure}

We rewrite the numerator of Eq.~\eqref{eq:def_g1} as
\begin{align}
\ev{\oa^\dagger(t) \oa(0)}_\mathrm{ss} = \Tr \bqty{\oa e^{\mathcal{L} t} (\oa \dosl^\mathrm{S,ss})}\,,
\end{align}
and evaluate this numerically.
As shown in Fig.~\ref{fig:g1_1},  
the temporal behavior qualitatively changes depending on $\lambda$.
The correlation function oscillates in time, and approaches zero
for a long time in the case of weak coupling.
This decay behavior is caused by dephasing from the system--reservoir coupling.
As $\lambda$ increases, the correlation function tends to remain positive for a longer time, which indicates that the coherence of the cavity mode is preserved for strong $\lambda$.

\begin{figure}
	\centering
	\includegraphics[width=3.4in]{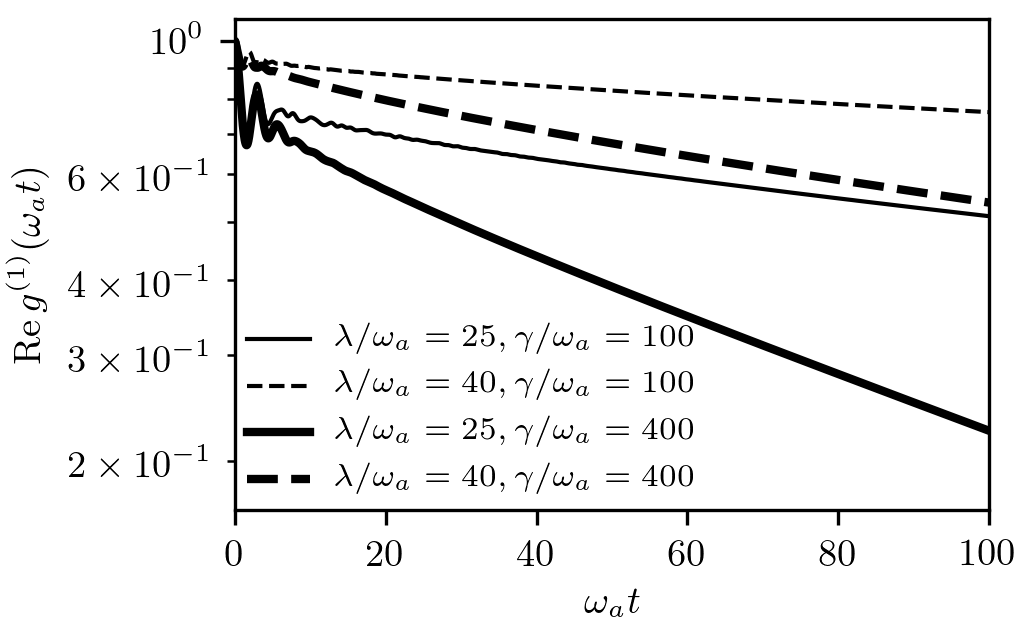}
	\caption{The first-order correlation function $\Re g^{(1)}$ on a logarithmic scale when $\lambda$ is significantly larger than $\lambda_c$ and $\gamma/\omegaa = 100, 400$.
		The other parameters are fixed as $N = 16$ and $\omegac / \omegaa = 400$.}
	\label{fig:competition}
\end{figure}

To study the competition between the preservation of the coherence 
and the reservoir-induced dephasing, we evaluate the correlation function 
in the deep superradiant region $\lambda \ge 2.5 \lambda_c$ 
for various values of $\lambda$ and $\gamma$.
The results are depicted in Fig.~\ref{fig:competition}.
We observe that $\Re g^{(1)}$ exhibits an almost exponential decay after a small oscillation, and $\Re g^{(1)}(t)$ is approximately parameterized
by $e^{-t/\tau_\mathrm{c}}$, where $\tau_\mathrm{c}$ is a coherence time.
As $\lambda$ becomes larger or the dephasing parameter $\gamma$ becomes
smaller, $\tau_\mathrm{c}$ becomes larger.

\begin{figure}
	\centering
	\includegraphics[width=3.4in]{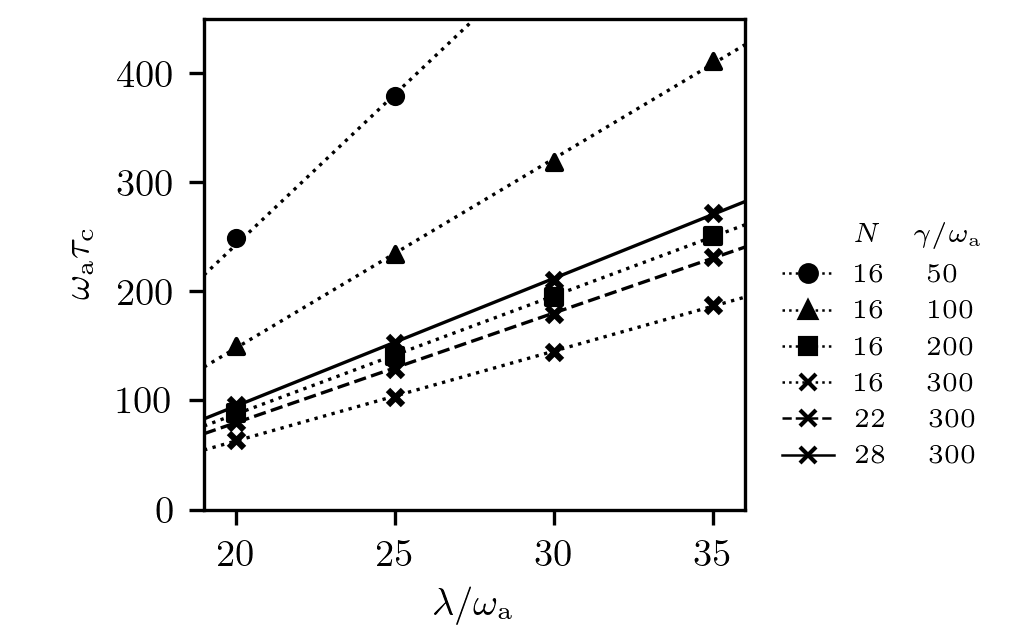}
	\caption{The points represent the coherence times obtained by fitting 
		$\Re g^{(1)}$
		by an exponential function. For fixed $N$ and $\gamma / \omegaa$, 
		the points lie on an auxiliary straight line. 
		The other parameter is fixed as $\omegac / \omegaa = 400$.}
	\label{fig:lambda_tauc}
\end{figure}

In order to capture this competition in a more quantitative manner, we fit $\Re g^{(1)}(t)$
by a function $C \exp (-t/\tau_c)$ with two parameters $C$ and $\tau_\mathrm{c}$ in
the decay region. The coherence times $\tau_c$ fitted in this manner
are depicted in Fig.~\ref{fig:lambda_tauc}.
The coherence time is approximately a linear function of
$\lambda$, which means that the coherence, or an ordered state, resulting from the atom--cavity interaction is enhanced to partly overbear dephasing in the superradiant region.
An increase in the number of atoms $N$ also enhances the coherence.

\section{Summary and outlook}
\label{sec:conclusion}

The dynamical properties of the finite-size Dicke model in a dispersive regime, coupled to a thermal reservoir with counter-rotating terms, have been investigated.
We have focused on the low-energy dynamics in the dispersive regime, where the low-energy sector description works well.
Then, the QME in the low-energy sector, which describes the dissipative dynamics of the Dicke system, is derived as Eq.~\eqref{eq:QME_simplified_Schrodinger}.
In the derivation of the QME, we assumed that the spectral density of the reservoir is given in an ohmic form, and that the reservoir temperature lies within the range presented in Eq.~\eqref{eq:beta_region}, in order to separate the time scales between the system and the reservoir.
Then, despite of the presence of the counter-rotating terms in the system-reservoir coupling, the QME becomes Markovian, which enables us to analyze the QME numerically without difficulty.

Using the QME, we have demonstrated that the steady state is uniquely determined as the maximum mixed state independently of the atom--cavity coupling.
The above implies that there is no bifurcation in the steady state, in contrast to the semi-classical description with the Lindblad dissipater \cite{Nagy2011,Oztop2012,Brennecke2013}.
On the other hand, we observed that the dumping form of $\Re \ev{\oa}$ qualitatively changes depending on the atom--cavity coupling strength, appearing as a long-lived state with $\Re \ev{\oa}$. Furthermore, this transition point approximately coincides with that of the superradiant transition determined in the thermodynamical limit of the isolated model.
Thus, we have found that the bifurcation in the Dicke system coupled to a thermal reservoir occurs in the long-lived state rather than in the steady state.

We have also calculated the first-order correlation function $g^{(1)}$ in the steady state.
The result shows that the cavity mode maintains coherence for a long time in the superradiant region of the atom--cavity coupling $\lambda$.
The stronger $\lambda$ becomes, the more the coherence in the superradiant region is enhanced to partly overcome dephasing.

We have assumed that the spectral density of the reservoir takes an ohmic form, in order to obtain computable expressions such as Eq.~\eqref{eq:C(t)_ohmic}, which help to capture the separation of the time scales.
For other spectral densities, further consideration would be desirable.

\appendix

\section{Correlation function in the ohmic dissipation}
\label{app:ohmic_cf}
We will evaluate the correlation function \eqref{eq:C(t)} with the ohmic spectral density $J(\omega) = \eta \omega e^{-\omega/\cutoff}$, where $\cutoff$ denotes a cutoff frequency, which will be taken to an infinite limit in the end.

First, we decompose $C(t)$ into two parts: $C(t) = C_\mathrm{th.} (t) + C_\mathrm{vac.} (t)$
\begin{align*}
C_\mathrm{th.} (t) &= 2 \int_0^\infty d \omega \, J(\omega) n_\mathrm{B}(\omega) \cos(\omega t) \,,\\
C_\mathrm{vac.} (t) &= \int_0^\infty d \omega \, J(\omega) e^{-i \omega t} \,.
\end{align*}
We introduce the formula
\begin{align}
\int_0^\infty d\omega \, \frac{\sin(\omega t)}{e^{\beta \omega} - 1}
=
\frac{\pi}{2 \beta} \coth \frac{\pi t}{\beta} - \frac{1}{2t}
\end{align}
for $t > 0$.
Differentiating this formula with respect to $t$, we obtain
\begin{align}
\int_0^\infty d\omega \, \frac{\omega \cos(\omega t)}{e^{\beta \omega} - 1}
=
- \frac{\pi^2}{2 \beta^2 \sinh^2 (\pi t / \beta)} + \frac{1}{2 t^2}\,.
\end{align}
Thus, we have that
\begin{align}
C_\mathrm{th.} (t) \approx -\frac{\eta \pi^2}{\beta^2 \sinh^2 (\pi t / \beta)} + \frac{\eta}{t^2} + \Order{\cutoff^{-1}}\,.
\label{eq:C_ohmic_th}
\end{align}
On the other hand, $C_\mathrm{vac.}$ is evaluated as
\begin{align}
C_\mathrm{vac.} (t) = \frac{-\eta}{(t - i \cutoff^{-1})^2}\,.
\label{eq:C_ohmic_vac}
\end{align}
From Eqs. \eqref{eq:C_ohmic_th} and \eqref{eq:C_ohmic_vac}, we have that
\begin{align}
C(t) &= -\frac{\eta \pi^2}{\beta^2 \sinh^2 (\pi t / \beta)} + \Order{\cutoff^{-1}} \,,
\end{align}
We eliminate the last term by taking the limit $\cutoff \to \infty$.
This result is valid only for $t > 0$.
Taking into account the symmetry $C(-t) = [C(t)]^\ast$, we obtain Eq.~\eqref{eq:C(t)_ohmic}.

\section{Evaluation of the dissipative term}
\label{app:Evaluation_diss}

Using Eqs.~\eqref{eq:def_L}, \eqref{eq:markov_approx}, and \eqref{eq:expansion_density_op}, a matrix element of the dissipative term in the QME Eq.~\eqref{eq:QME_Born} is
\begin{widetext}
	\begin{multline}
	\bra*{\phi_\mu} \mathcal{L}_\mathrm{R} (t,t-s) \dosl \ket*{\phi_\nu}
	= \sum_{\mu'= -J}^J \int d\omega \, \bigg\{ \Big[ G^{(2)}_{\mu,\mu'} (t,t-s) \dosl_{\mu',\nu}
	- G^{(1)}_{\mu, \mu'} (t-s) \dosl_{\mu',\nu} G^{(1)}_{\mu, \mu'} (t) \Big] \tilde{C}(\omega) e^{-i \omega s} \\
	+ \Big[ \dosl_{\mu,\mu'} G^{(2)}_{\mu',\nu} (t-s,t)
	- G^{(1)}_{\mu, \mu'} (t) \dosl_{\mu',\nu} G^{(1)}_{\mu, \mu'} (t-s) \Big] \tilde{C}(\omega) e^{i \omega s} \bigg\}\,,
	\label{eq:mel_L}
	\end{multline}
\end{widetext}
where $G^{(1)}$ and $G^{(2)}$ are defined by $G^{(1)}_{\mu,\nu} (t) = \bra*{\phi_\mu} \oS(t) \ket*{\phi_\nu}$ and $G^{(2)}_{\mu,\nu} (t_1,t_2) = \bra*{\phi_\mu} \oS(t_1) \oS(t_2) \ket*{\phi_\nu}$, respectively.
Here, we suppress the variable $t$ of $\dosl(t)$. Now,
$\oS(t)$ has a complicated time dependence, because $\oH_0$ includes the interaction between the atom and the cavity mode.
Then, using the projection operator $\Pi_\mu = \dyad{\mu}_\text{atom} \otimes 1_\text{cavity}$, where $\ket{\mu}$ is an eigenstate of $\oJ{1}$ with the eigenvalue $\mu$, we can express $\oa(t)$ in the following brief form:
\begin{align}
\Pi_\mu \oa(t) \Pi_\mu = \pqty{\oa - \alpha_\mu} e^{-i \omegac t} + \alpha_\mu\,,  
\end{align}
where $\alpha_\mu$ is defined in Eq.~\eqref{eq:def_D_alpha}.
Computing straightforwardly, we have that
\begin{align}
\begin{aligned}
G^{(1)}_{\mu,\nu} (t) &= - \delta_{\mu,\nu} (d+d^\ast) \frac{2 \lambda \mu}{\sqrt{N} \omegac} \,, \\
G^{(2)}_{\mu,\nu} (t_1,t_2) &= \delta_{\mu,\nu} \bqty{ \abs{d}^2 e^{-i \omegac (t_1-t_2)} + (d + d^\ast)^2 \frac{4 \lambda^2 \mu^2}{N \omegac^2}} \,.
\end{aligned}
\label{eq:def_G}
\end{align}
Substituting Eq.~\eqref{eq:def_G} into Eq.~\eqref{eq:mel_L} and integrating over $s$ from $0$ to $\infty$, we obtain 
\begin{multline}
\dosl_{\mu,\nu} \bigg[
2 \tilde{C}(-\omegac) \abs{d}^2 + \tilde{C}(0) (d+d^\ast)^2 \frac{4 \lambda^2}{N \omegac^2} (\mu - \nu)^2
\\- 
i \mathrm{PV} \int d \omega \, \frac{\tilde{C}(\omega)}{\omega} (d+d^\ast)^2 \frac{4 \lambda^2}{N \omegac^2} (\mu^2 - \nu^2)
\bigg]\,,
\end{multline}
which corresponds to
$-i\comm{\Sigma}{\dosl}_{\mu,\nu} - [\Gamma^\text{ex.}   + \Gamma^\text{de.}_{\mu,\nu}] \dosl_{\mu,\nu}$ in Eq.~\eqref{eq:QME_full}.



%

\end{document}